\documentclass[twocolumn]{aastex631}

\usepackage{float}
\usepackage{graphicx}

\usepackage[switch]{lineno}

\begin{document}
\raggedbottom


\title{GTC/CanariCam Mid-IR Polarimetry of Magnetic Fields in Star-Forming Region W51 IRS2}


\author{Charles M. Telesco}
    \affiliation{Department of Astronomy, University of Florida, 211 Bryant Space Science Center, FL 32611, USA}

\author{Han Zhang}
    \affiliation{Department of Astronomy, University of Florida, 211 Bryant Space Science Center, FL 32611, USA}

\author{Frank Varosi}
    \affiliation{Department of Astronomy, University of Florida, 211 Bryant Space Science Center, FL 32611, USA}
    
\author{Pae Swanson}
    \affiliation{Department of Astronomy, University of Florida, 211 Bryant Space Science Center, FL 32611, USA}

\author{Sergio José Fernández Acosta}
    \affiliation{Gran Telescopio de Canarias, Cuesta de San José s/n, E-38712, Breña Baja, La Palma, Spain}

\author{Christopher M. Wright}
    \affiliation{School of Science, UNSW Canberra, PO Box 7916, Canberra BC 2610, Australia}

\author{Christopher Packham}
    \affiliation{Physics and Astronomy Department,
    University of Texas at San Antonio, San Antonio, TX 78249, USA}


\begin{abstract}
We present 0\farcs4-resolution imaging polarimetry at 8.7, 10.3, and 12.5 $\mu$m, obtained with CanariCam at the Gran Telescopio Canarias (GTC), of the central 0.11 pc x 0.28 pc (4\farcs2 x 10\farcs8) region of W51 IRS2. The polarization, as high as $\sim$14\%, arises from silicate particles aligned by the interstellar magnetic field (B-field). We separate, or unfold, the polarization of each sightline into emission and absorption components, from which we infer the morphologies of the corresponding projected B-fields that thread the emitting and foreground-absorbing regions. We conclude that the projected B-field in the foreground material is part of the larger-scale ambient field. The morphology of the projected B-field in the mid-IR emitting region spanning the cometary \ion{H}{2} region W51 IRS2W is similar to that in the absorbing region. Elsewhere, the two B-fields differ significantly with no clear relationship between them. The B-field across the W51 IRS2W cometary core appears to be an integral part of a champagne outflow of gas originating in the core and dominating the energetics there. The bipolar outflow, W51north jet, that appears to originate at or near SMA1/N1 coincides almost exactly with a clearly demarcated north-south swath of lower polarization. While speculative, comparison of mid-IR and submm polarimetry on two different scales may  support a picture in which SMA1/N1 plays a major role in the B-field structure across W51 IRS2.   
\end{abstract}


\keywords{ISM -- 
        IR polarimetry -- 
        magnetic fields -- 
        W51 -- 
        cometary outflows}

\section{Introduction} \label{sec:intro}

The W51 giant molecular cloud is one of the most active and massive star-forming regions in the Galaxy. It is located at a distance of 5-8 kpc and contains an aggregation of \ion{H}{2} regions and young stellar objects evident as the major complexes A, B, and C (Ginsburg et al. \hyperlink{gin15}{2015}, \hyperlink{gin17}{2017}). Besides being an exceptionally rich environment for the study of the evolution of massive stars, W51 exhibits little foreground and background contamination, further enhancing its utility for our polarimetric study and many others. The W51 A complex contains two protocluster regions, W51 main and IRS2. Located at approximately 5.4 kpc (Xu et al. \hyperlink{xu09}{2009}) and the focus of the current study, W51 IRS2 consists of many \ion{H}{2} regions and YSOs, some of which are massive ($\geq$50 $M_\odot$) stellar candidates (Barbosa et al. \hyperlink{barbosa08}{2008}).
    
W51 IRS2 has been studied at multiple wavelengths spanning the near-infrared (NIR) to radio (e.g., Tang et al. \hyperlink{tang13}{2013}; Zapata et al. \hyperlink{zap09}{2009}). The most prominent sources in this region at mid-infrared (mid-IR) wavelengths are IRS2E and IRS2W (Figure 1; Barbosa et al. \hyperlink{barbosa16}{2016}; Okamoto et al. \hyperlink{oka01}{2001}). Detection of the NIR CO overtone in IRS2E suggests the presence of an accretion disk buried deeply within this heavily extinguished source (A\textsubscript{V} $\simeq$ 63 mag; Barbosa et al. \hyperlink{barbosa08}{2008}, \hyperlink{barbosa16}{2016}). IRS2W is an ultra-compact \ion{H}{2}  (UC \ion{H}{2}) region with a comet-shaped core  and an O3 ionizing star (Barbosa et al. \hyperlink{barbosa08}{2008}). Mid-IR sources also include a high velocity, blue-shifted jet (the Lacy jet) first discovered in the W51 IRS2 region by Lacy et al. (\hyperlink{lacy07}{2007}) from [\ion{Ne}{2}] and [\ion{S}{4}] line emission; it appears to be a molecular outflow ionized as it emerges into the \ion{H}{2} region from a source to the west and outside our field of view (Ginsburg et al. \hyperlink{gin17}{2017}). 

W51 IRS2 offers the opportunity to study the magnetic field (B-field) in a complex star forming environment that is close enough to permit us to usefully disentangle the complex morphologies associated with different sources at a range of evolutionary stages. Dust polarization is commonly used to trace B-fields in star formation regions. Non-spherical dust grains tend to spin, perhaps as a result of radiative alignment torques from an anisotropic radiation field (Lazarian \hyperlink{laz07}{2007}; Cho \& Lazarian \hyperlink{cho07}{2007}), and become preferentially aligned with respect to a local B-field. This alignment results in polarized extinction by the dust of background starlight transmitted through the ensemble of particles. That transmitted light is polarized parallel to the B-field, whereas thermal emission from the ensemble of particles is polarized perpendicular to the B-field lines. Therefore, provided that one can distinguish between polarization resulting from emission and absorption by dust, polarimetry can be used to infer the projected B-field directions.

\begin{figure}
\resizebox{\hsize}{!}{\includegraphics{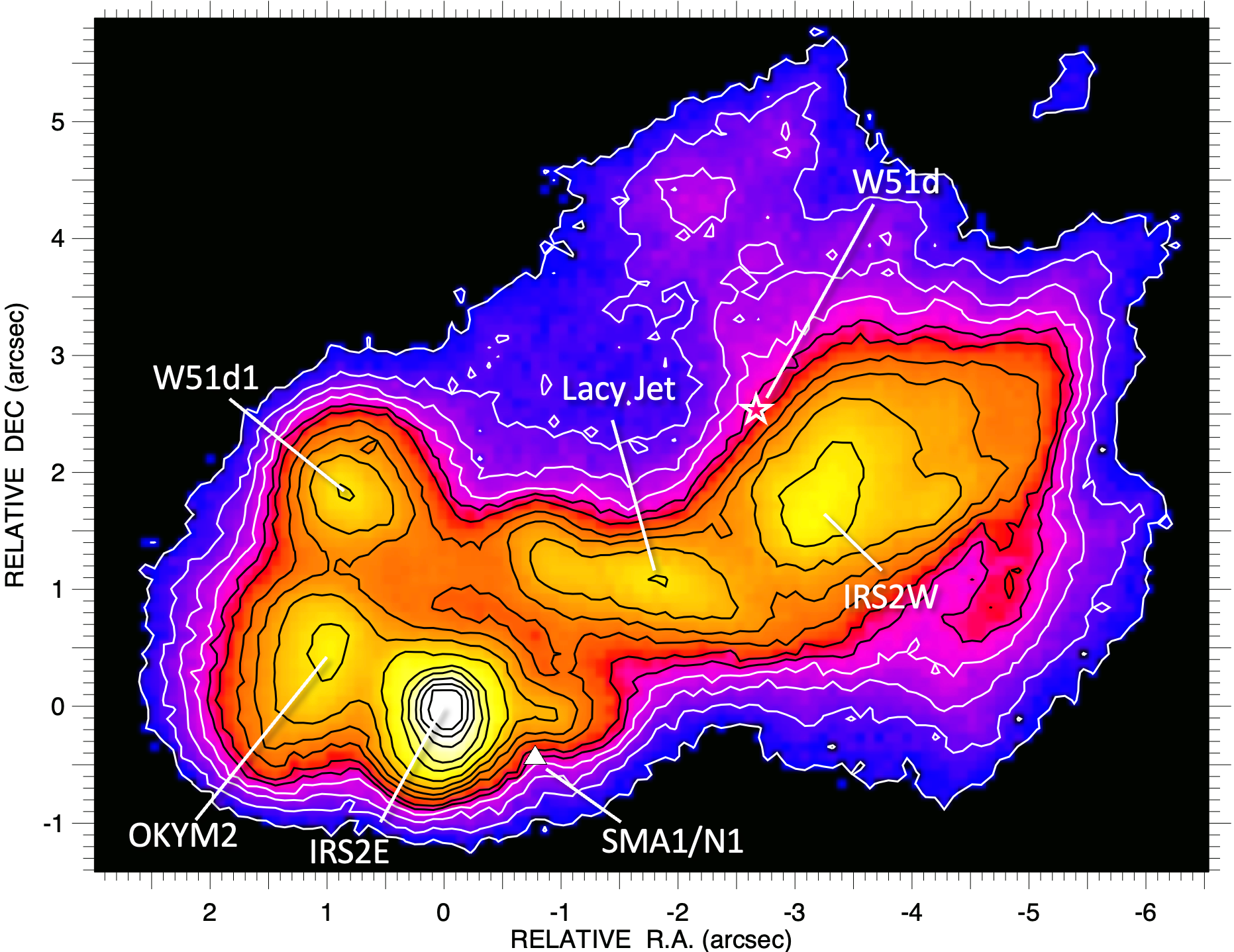}}%
\caption{
    12.5 $\mu$m CanariCam image of W51 IRS2 indicating principal mid-IR sources (this work; see also Figuerêdo et al. \protect\hyperlink{fig08}{2008}; Barbosa et al. \protect\hyperlink{barbosa16}{2016}). Star symbol indicates approximate position of exciting star assumed coincident with near-IR peak W51d (Barbosa et al. \protect\hyperlink{barbosa08}{2008}, \protect\hyperlink{barbosa16}{2016}, Okamoto et al. \protect\hyperlink{oka01}{2001}). Triangle symbol locates SMA1 (Tang et al. \protect\hyperlink{tang13}{2013}), which is 0\farcs08 from N1 (Koch et al. \protect\hyperlink{koch18}{2018}) assumed coincident with SMA1. Intensity contours, as percentage of top contour level, are: 3, 4, 5, 6, 7, 8, 10, 12, 15, 20, 30, 40, 60, 80. Top contour corresponds to 100 Jy per square arcsec. Image is not smoothed and results from co-added, on-source integration time of 66 s. RA and Dec offsets are from IRS2E located at 19\textsuperscript{h}23\textsuperscript{m}40.10\textsuperscript{s}, 14\degr 31\arcmin 06.0\arcsec (J2000) (Lacy et al. \protect\hyperlink{lacy07}{2007}).
}\label{Figure1}%
\end{figure}

Non-magnetic effects can also produce polarization. The mechanisms include scattering from dust grains in an anisotropic radiation field (Kataoka et al. \hyperlink{kat15}{2015}; Yang et al \hyperlink{yang16}{2016}) and dust alignment along the direction of the radiation field (Tazaki et al. \hyperlink{taz17}{2017}). Theory and observations (Stephens et al. \hyperlink{steph17}{2017}; Kataoka et al. \hyperlink{kat16}{2016}) indicate that, under some circumstances, these processes may contribute to the observed polarization at particular wavelengths, but they are not considered to be relevant here.

We present mid-IR polarimetric imaging observations of W51 IRS2 with an angular resolution of $\sim$0\farcs4 (0.011 pc). The field of view centered on IRS2 is 0.11 pc $\times$ 0.28 pc. In addition to exploring the projected B-field morphology, we use the method of ``polarimetric tomography'' to infer spatial structure of the B-field along the sightline. This approach was explored in detail by Barnes et al. (\hyperlink{barnes15}{2015}) for the star-forming region K3-50. These data complement the millimeter and submillimeter polarimetry obtained by Tang et al. (\hyperlink{tang13}{2013}), Koch et al. (\hyperlink{koch18}{2018}), and Chrysostomou et al. (\hyperlink{chr02}{2002}) of W51 IRS2 and the surrounding regions. In general, due to opacity and/or temperature effects, mid-IR emission traces different regions than does mm/submm emission. 

The paper is organized as follows: in \hyperref[sec:2]{Section 2}, we describe the mid-IR observations and data reduction of W51 IRS2; in \hyperref[sec:3]{Section 3}, we summarize the Aitken method for separating, or unfolding, the polarization due to absorption and emission along a sightline, apply it to prominent sources in W51 IRS2, and consider general trends in the B-field distribution across this region; in \hyperref[sec:4]{Section 4} we examine in more detail the implications of the inferred B-field structure in the absorbing and emitting regions in the context of gas outflows and accretion; and in \hyperref[sec:5]{Section 5}, we summarize our conclusions.

\section{Observations \& Data Reduction} \label{sec:2}

We obtained the polarimetric images of W51 IRS2 using three mid-IR filters (Si2-8.7 $\mu$m, Si4-10.3 $\mu$m, and Si6-12.5 $\mu$m) with CanariCam in 2013 August and 2014 June as part of the CanariCam Science Team program (Table 1). CanariCam is the mid-IR multimode (imaging, spectroscopy, and polarimetry) facility camera on the 10.4 m Gran Telescopio CANARIAS (GTC) on La Palma, Spain (Telesco et al. \hyperlink{telesco03}{2003}). It has a field of view of 26\arcsec $\times$ 19\arcsec with a pixel scale of 0\farcs079. Polarimetry is accomplished through the use of a Wollaston prism, which produces a separation between the \emph{o} and \emph{e} (ordinary and extraordinary) rays, and a half-wave plate rotated sequentially to four orientations (0\degr, 45\degr, 22.5\degr, and 67.5\degr) during observations.

\begin{figure*}[t]
\centering\includegraphics[width=\textwidth]{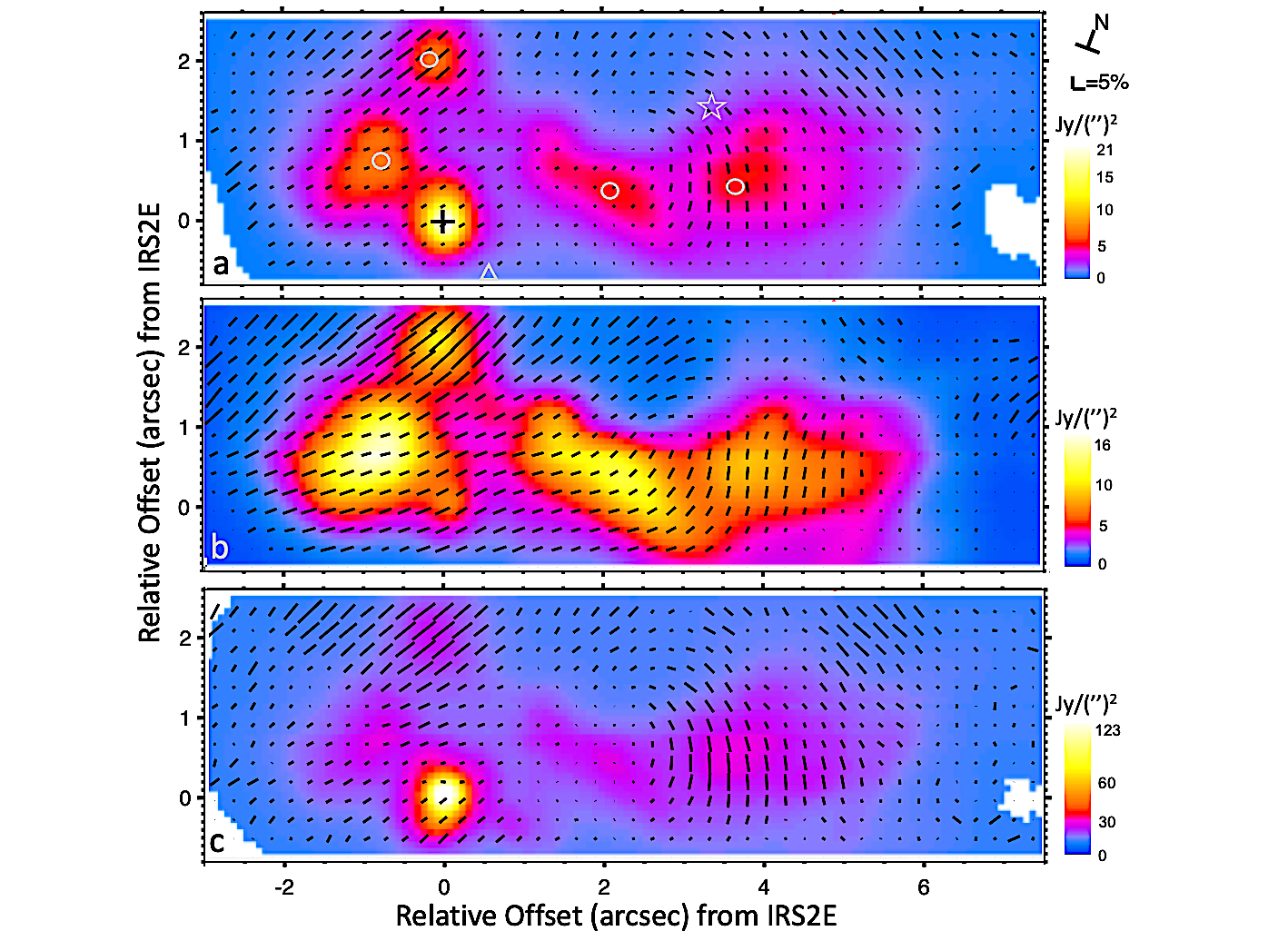}
\caption{
   Polarization across W51 IRS2 at (a) 8.7 $\mu$m, (b) 10.3 $\mu$m, and (c) 12.5 $\mu$m. Colored background in each image is Stokes I (total intensity in Jy per square arcsec) at that wavelength. Tilted line segments plotted at bin centers correspond to observed polarization magnitudes \emph{p}(\%)  and orientations. Data have been smoothed by 3 $\times$ 3-pixel (0\farcs24 $\times$ 0\farcs24) binning. Measurements are shown only at locations where Stokes I SNR $>$ 10. Symbols in top panel indicate locations of named sources in Figure 1, with the cross denoting the position of IRS2E. Note 20\degr tilt of field with respect to north.
}\label{Figure2}
\end{figure*}

The polarimetric imaging observations of W51 IRS2 were interlaced with observations of the standard star Vega for flux and point-spread-function calibration (Cohen et al. \hyperlink{cohen99}{1999}) and the standard star AFGL 2591 from Smith et al. (\hyperlink{smith00}{2000}) to calibrate the polarization position angle (PA). Standard chop-nod sequences were used with a chop-throw of 15\arcsec at PA = 20\degr\ to minimize contamination from background emission. The CanariCam field of view was also rotated by 20\degr\ from the north (i.e., the short, or vertical, axis of the image is along PA = 20\degr) to enable more optimal sampling with the polarimetry mask of the IRS2 field. In the polarimetry mode, the actual field of view is reduced to 26\arcsec x 2\farcs6 by insertion of a focal mask to avoid overlapping between \emph{o} and \emph{e} rays originating simultaneously at different source locations. Since the W51 IRS2 emission region of interest is larger than a single field of view in the polarization mode, we split our observations of the target into two parts offset from each other but overlapping by a few pixels for registration and to avoid edge effects. 
    
A CanariCam 10.3 $\mu$m image of the same field shown in Figure 1 at 12.5 $\mu$m served as a guide to construct the final mosaic of the polarization map from two polarization mask slots and telescope pointings; the relative positions of W51d1 and IRS2E were used to register and build the mosaic of the region. The achieved angular resolution (full width at half-maximum intensity) for the polarimetric imaging was 0\farcs35-0\farcs40 (Table 1), slightly worse than the mid-IR diffraction limit of 0\farcs3. 
    
The data were reduced with custom IDL software (Li \hyperlink{li14}{2014}; F. Varosi, unpublished). We computed normalized Stokes parameters $q$ and $u$, where $q = Q/I$ and $u = U/I$. The degree of polarization is $p = (q\textsuperscript{2} + u\textsuperscript{2})\textsuperscript{0.5}$ and the polarization position angle is computed as PA = $0.5\arctan(u/q)$. The uncertainties $\sigma_{q}$ and $\sigma_{u}$ associated with the normalized Stokes parameters were assumed to equal 1/SNR, where SNR is the signal-to-noise ratio at each corresponding pixel, generally found to be a conservative good estimate. The uncertainties were then propagated through the computation to obtain $\sigma_{p}$ and the PA uncertainty $\sigma$$\textsubscript{PA}$ = $\sigma_{p}/2$. The degree of polarization is then debiased using $\sigma_{p}$ with the modified asymptotic estimator developed by Plaszczynski et al. (\hyperlink{pla14}{2014}). Instrumental polarization, which is typically $\sim$0.6\%, was determined during the commissioning of CanariCam, with subsequent observations indicating that its behavior is stable and well characterized. We corrected the instrumental polarization in the $Q-U$ plane. 
Aperture polarization measurements of individual sources in this region are given in Table 2, with the size of the aperture taken to be 0\farcs4 $\times$ 0\farcs4 centered on source. Repeated measurements of photometric and PA calibrators, particularly AFGL 2591, indicate systematic uncertainties of order $\pm$10\% in the photometry, $\pm$0.2\% in polarization, and $\pm$5$\degr$ in PA. These systematic uncertainties are not shown explicitly in Table 2. Note that systematic photometric uncertainties do not contribute to uncertainties in the polarimetry.

The polarization images obtained in the three passbands are presented in Figure 2. The colored background corresponds to the total observed intensity (Stokes I) distribution at each wavelength, and the line-segment length and orientation indicate the fractional polarization \emph{p}(\%) and its PA at each point. Data are only plotted where SNR $>$ 10 for Stokes I. These displayed data have been smoothed by 3 $\times$ 3-pixel (0\farcs24 $\times$ 0\farcs24) binning.
 
\section{Results} \label{sec:3}

{\subsection{Polarization Components: The Aitken Method}} \label{sec:3.1}
    
The observed mid-IR polarization can be explained as arising from thermal emission and/or absorption from aligned non-spherical dust grains. Unique to the mid-IR, both processes may apply simultaneously along a given sightline, such as when an emitting source is embedded in an absorbing cocoon, with each region threaded by its own B-field. Strictly speaking, radiative transfer within the entire region should be analyzed self-consistently to interpret the observations. However, Aitken et al. (\hyperlink{ait04}{2004}) developed a relatively simple multiwavelength method to separate, or unfold, the emissive and absorptive polarization components using template spectral signatures of silicates in the Orion nebula. The Aitken method permits a separation of the observed polarization into two contributions: the polarization originating from aligned dust in the mid-IR emitting region and the polarization originating from aligned dust in a foreground absorbing region. We provide more information about the Aitken method in the \hyperlink{appendix}{Appendix}.

\vspace{-.65in} 

\begin{deluxetable*}{cccccc}\label{table1}
\tablenum{1}\caption{Observation Log}
    \tablehead{\colhead{Imaging} & \colhead{Filters} & \colhead{$\Delta\lambda$} & \colhead{Date} & \colhead{Integration} & \colhead{FWHM (PSF)} \\ 
    \colhead{} & \colhead{($\mu$m)} & \colhead{($\mu$m)} & \colhead{(UT)} & \colhead{Time (s)} & \colhead{($\arcsec$)}
} 
\startdata
    & Si2(8.7) & 1.1 & 2014 June 18 & 509 & 0.40 \\
    & Si4(10.3) & 0.9 & 2014 June 18 & 509 & 0.35 \\
    & Si6(12.5) & 0.7 & 2014 June 9 & 761 & 0.40 \\
    \hline
    & Si2(8.7) & 1.1 & 2013 Aug 20 & 582 & 0.40 \\
    & Si4(10.3) & 0.9 & 2013 Aug 20 & 582 & 0.38 \\
    & Si6(12.5) & 0.7 & 2013 Sep 1 & 761 & 0.36 \\
\enddata
\end{deluxetable*}

\begin{deluxetable*}{ccccc}\label{table2}
\tablenum{2}\caption{Polarization and Flux Measurements}
    \tablehead{\colhead{Object} & \colhead{$\lambda$} & \colhead{Flux Density} & \colhead{\emph{p}} & \colhead{PA} \\ 
    \colhead{} & \colhead{($\mu$m)} & \colhead{(Jy)} & \colhead{(\%)} & \colhead{($\degr$)}
} 
\startdata
    W51d1 & 8.7  & 0.98 & 7.3 (0.2) & 150 (1) \\
          & 10.3 & 1.42 & 13.4 (0.2) & 150 (1) \\
          & 12.5 & 1.87 & 11.7 (0.2) & 149 (1) \\
    \hline
    OKYM2 & 8.7  & 1.04 & 2.9 (0.1) & 129 (1) \\
          & 10.3 & 2.24 & 4.0 (0.1) & 127 (1) \\
          & 12.5 & 2.21 & 3.1 (0.2) & 126 (1) \\
    \hline
    IRS2E & 8.7  & 2.43 & 2.5 (0.1) & 141 (1) \\
          & 10.3 & 0.92 & 5.6 (0.3) & 139 (2) \\
          & 12.5 & 13.07 & 2.3 (0.1) & 140 (1) \\
    \hline
    Lacy Jet & 8.7  & 0.78 & 1.6 (0.2) & 162 (4) \\
             & 10.3 & 1.76 & 3.5 (0.2) & 146 (1) \\
             & 12.5 & 2.12 & 0.5 (0.2) & 172 (11) \\
                \hline
    W51d & 8.7  & 0.24 & 2.9 (0.7) & 77 (7) \\
         & 10.3 & 0.27 & 2.1 (0.8) & 115 (11) \\
         & 12.5 & 0.69 & 4.9 (0.6) & 64 (4) \\
    \hline
    IRS2W & 8.7  & 0.80 & 4.0 (0.2) &  30 (2) \\
          & 10.3 & 1.28 & 4.7 (0.2) &  14 (1) \\
          & 12.5 & 2.51 & 6.6 (0.2) &  32 (1) \\
\enddata
\tablecomments{
   Flux densities are totals within 0\farcs4 $\times$ 0\farcs4 bin centered on source. All flux densities were measured with signal-to-noise ratios (SNRs) greater than 200 except for W51d, for which SNR $\approx$ 90. Non-systematic measurement errors in flux densities are therefore negligible, with their main uncertainty being $\pm$10\% due to flux calibration. For \emph{p} and PA, tabulated uncertainties reflect only measurement statistics; systematic uncertainties, which are not tabulated, are $\pm$0.2$\%$ for \emph{p} and $\pm$5$\degr$ for PA. All position angles correspond to east from north.
}

\end{deluxetable*}

The effectiveness of the Aitken method relies on the fact that the mid-IR polarization efficiency as a function of wavelength differs for the silicate spectral feature depending on whether the radiation is transmitted through, or emitted by, a population of mutually aligned non-spherical silicate particles. Polarization measurements for at least two wavelengths are needed with this approach; we use observations at three wavelengths, one of which is near the center of the silicate feature, with the other two being near the long- and short-wavelength ends of the feature accessible through the 10 $\mu$m atmospheric window. This method is optimized to minimize $\chi$\textsuperscript{2} of the fit to the normalized Stokes parameters $q$ and $u$ independently. A principal assumption when using this technique is that the dust grains dominating the polarization are silicates. Based on consistency with observations of other sources and at other wavelengths, it works well for a variety of astronomical objects (e.g., Barnes et al. \hyperlink{barnes15}{2015}; Lopez-Rodriguez et al. \hyperlink{loro17}{2017}; Li et al. \hyperlink{li18}{2018}).
        
We apply Aitken's approach to our multiwavelength imaging polarimetry data, computing the decomposition for each 3 $\times$ 3-pixel (0\farcs24 $\times$ 0\farcs24) bin to increase the SNR and more closely match the angular resolution. Along each sightline, the least-squares fitting of the template polarization profiles to the observed multiwavelength polarization then permits determination of the relative contribution of the aligned grains located in the emitting and foreground absorbing regions at each wavelength $\lambda$. We thereby derive the distributions of the absorptive and emissive polarizations $p_{\text{a}}{(\lambda)}$ and $p_{\text{e}}{(\lambda)}$ with their corresponding PA\textsubscript{a} and PA\textsubscript{e} distributions.

Taking the classical dust alignment theory (e.g., Lazarian \hyperlink{laz07}{2007}; Draine \& Hensley \hyperlink{dra21}{2021}), elongated dust grains in the presence of a B-field spin with their angular momentum and grain minor axes tending to be aligned with the field direction $\theta$. Therefore, the differential extinction by an ensemble of such dust grains polarizes background starlight with direction parallel with the B-field ($\theta$\textsubscript{a} = PA\textsubscript{a}). In contrast, thermal emission from these dust grains is polarized in the direction perpendicular to the B-field; the field direction for this (emissive) polarization can therefore be inferred by rotating its observed PA by 90\degr \mbox{($\theta$\textsubscript{e} = PA\textsubscript{e} + 90\degr)}. We refer to the polarization line segments corresponding to the inferred B-fields projected on the plane of the sky and in the emitting and foreground absorbing regions as B\textsubscript{e} and B\textsubscript{a}, respectively. 
        
In Figure 3, which illustrates the approach, we show the decomposition of the emissive and absorptive polarization components at six locations (Figure 1) across W51 IRS2. For the relatively pure cases of W51d1 and OKYM2 in which emission dominates and the pure case of IRS2E in which absorption dominates, we see the expected constancy of the PA values across the observed wavelengths. For the other sources, the evident variation of PA with wavelength reflects the corresponding differences in the relative contributions of emissive and absorptive polarization components along those sightlines. Recall that, for polarization, PA values differing by 180\degr are equivalent. In the next section we discuss these and other individual sources. 

\begin{figure}
\resizebox{\hsize}{!}{\includegraphics{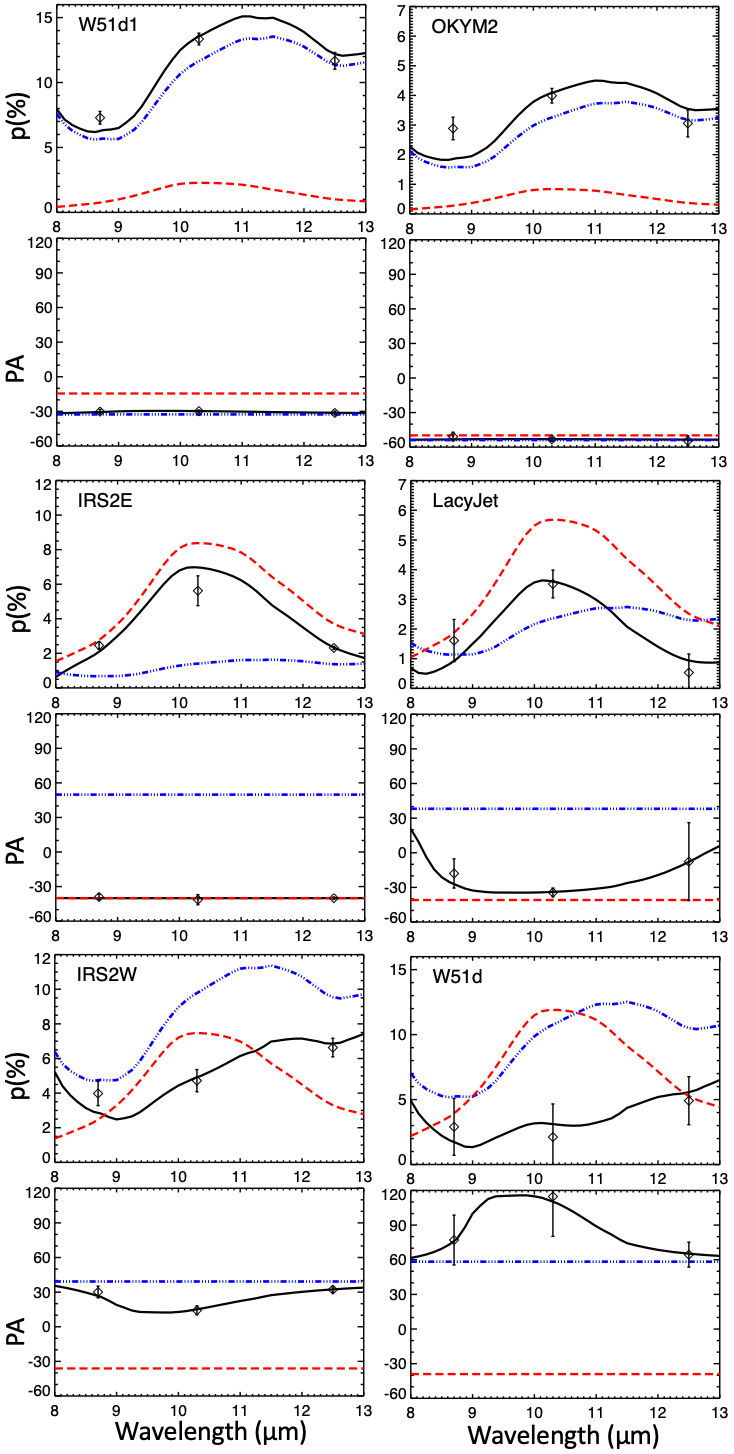}}
\caption{
    Decomposition of observed multiwavelength polarimetry (diamonds) of select sources in W51 IRS2 into absorptive (red line) and emissive (blue line) components. The black solid line is the composite fitting result. For each source, the lower panel shows the corresponding PA measured east from north.  
}\label{Figure3}
\end{figure}

In Figure 4 we show our derived line segments and colored background contours representing the unfolded spatial distributions of fractional polarizations $\emph{p}$(\%) at 10.3 $\mu$m, with the polarization line segments for the emitting region rotated by 90\degr. These plots reveal the morphologies of the projected B-fields in the absorbing and emitting regions of W51 IRS2, discussed in detail below.

\begin{figure*}
\includegraphics[width=\textwidth]{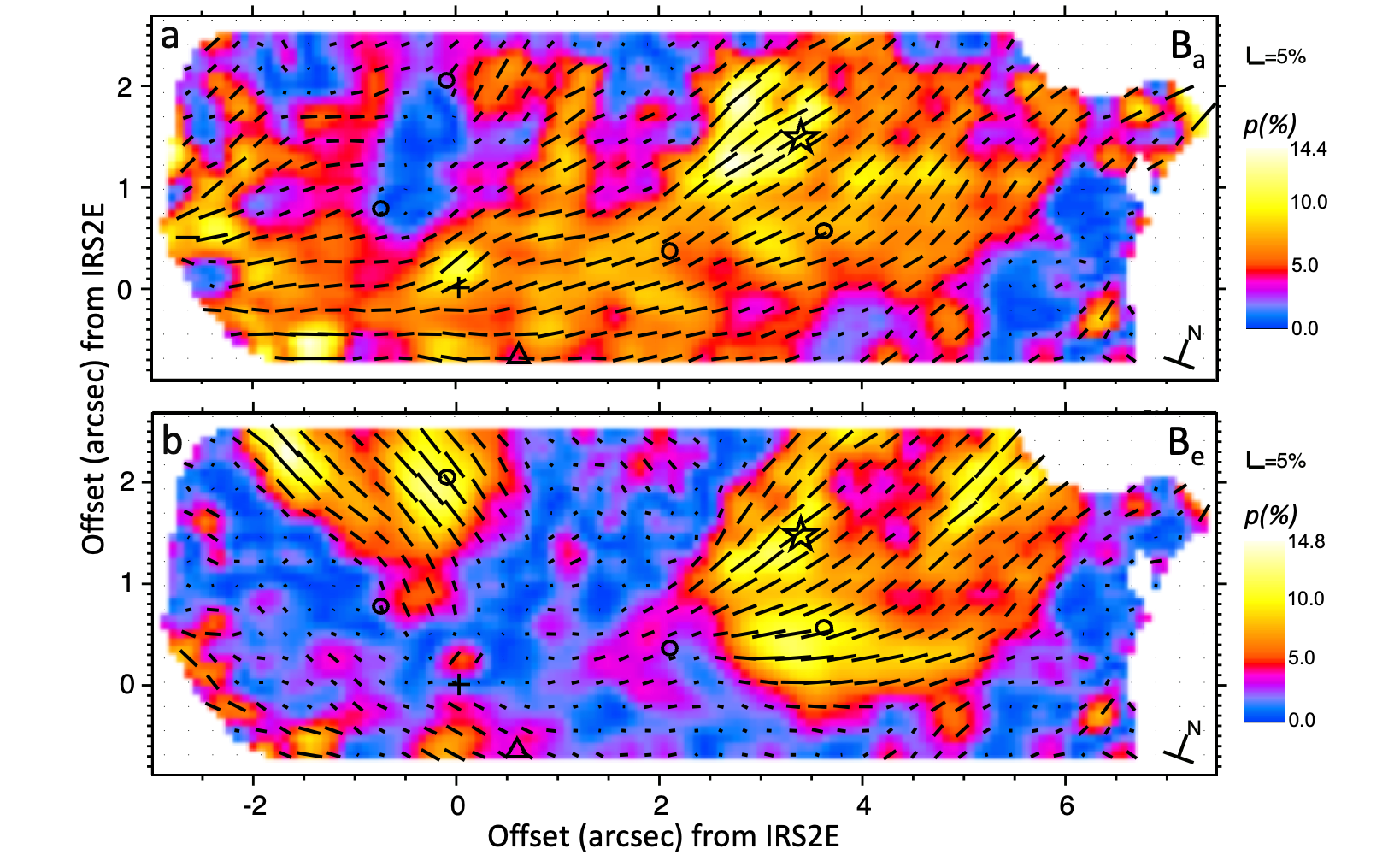}
\caption{
    Derived polarization at 10.3 $\mu$m in (a) the absorbing and (b) emitting regions, with line segments for the latter rotated by 90\degr. Line segment lengths are proportional to the fractional polarization $\emph{p}$(\%), and line segment orientations indicate corresponding directions of projected B-field lines. Colored background contours (scales at right) indicate derived fractional polarization $\emph{p}$(\%). Data have been smoothed by 3 $\times$ 3-pixel (0\farcs24 $\times$ 0\farcs24) binning. Locations are indicated for sources shown in Figure 1, including IRS2E (cross), SMA1/N1 (triangle), W51d (star).
}\label{Figure4}
\end{figure*}

{\subsection{Polarization of Individual Sources}}\label{sec:3.2}

W51 IRS2 contains multiple \ion{H}{2} regions and OB stars. Our observations resolve individual sources in this region: W51d1, OKYM2, IRS2E, the Lacy Jet, IRS2W, and W51d, with their corresponding locations, polarization measurements, and spectropolarimetric decompositions indicated in Figure 1, Table 2, and \mbox{Figure 3}, respectively. As we comment on each source, it is worth keeping in mind that the B-fields we derive for that particular sightline may be more indicative of large-scale trends spanning that region than of conditions local to that source. Our sub-arcsec imaging polarimetry provides a firm basis for distinguishing those larger trends from possible source-specific properties. 
\emph{W51d1 (OKYM5)} is seen as a NIR, mid-IR, and radio source. It is a UC \ion{H}{2} region and contains a compact cluster of stars (Figuerêdo et al. \hyperlink{fig08}{2008}). The polarized mid-IR radiation along this sightline arises principally from emitting dust. We infer $\theta$\textsubscript{e} $\simeq$ 60\degr for W51d1.
    
\emph{OKYM2} was identified by Okamoto et al. (\hyperlink{oka01}{2001}). The radio source appears to have a small offset (0\farcs2 - 0\farcs3) from the mid-IR emission peak. Located in a region of relatively low extinction (Barbosa et al. \hyperlink{barbosa16}{2016}), its polarization is dominated by the emissive component at PA $\simeq$ 127\degr, which corresponds to $\theta$\textsubscript{e} $\simeq$ 37\degr for the projected B-field.
    
\emph{IRS2E (OKYM1)} exhibits a deep silicate absorption feature (Okamoto et al. \hyperlink{oka01}{2001}) corresponding to \mbox{A\textsubscript{v} $\simeq$ 63 mag}, much higher than the average extinction in this region (Barbosa et al. \hyperlink{barbosa08}{2008}, \hyperlink{barbosa16}{2016}). Its NIR CO overtone emission indicates the presence of accretion in the inner disk (Barbosa et al. \hyperlink{barbosa08}{2008}). It is not known to be associated with a UC \ion{H}{2} region nor specifically with any of the 350 $\mu$m or 850 $\mu$m emission observed across this region with 20\arcsec-resolution by Dotson et al. (\hyperlink{dot10}{2010}) and Matthews et al. (\hyperlink{mat09}{2009}), respectively. IRS2E is the most embedded mid-IR source in the region and is likely a very young massive star probably still in the hot core phase transitioning to a UC \ion{H}{2} region (Zapata et al. \hyperlink{zap09}{2009}). Analysis by Okamoto et al. (\hyperlink{oka01}{2001}) suggests that it is located behind the ionized gas of the cluster and that the extinction is at least partly intrinsic to the embedded source. The picture is consistent with our polarimetry, which is fitted well with the absorptive component at \mbox{PA $\simeq$ 140\degr}.
\newpage
\emph{The Lacy jet} was first reported by Lacy et al. (\hyperlink{lacy07}{2007}) using the mid-IR [\ion{Ne}{2}] emission line and later confirmed with the H77$\alpha$ radio recombination line (Ginsburg et al. \hyperlink{gin16}{2016}). The CO counterpart of the jet was subsequently detected by Ginsburg et al. (\hyperlink{gin17}{2017}), who identified the likely base of the jet with the mm source ALMAmm31, which lies off to the SW of our image boundary. Ginsburg et al. (\hyperlink{gin17}{2017}) propose the scenario in which a molecular outflow becomes ionized when it emerges from the primarily neutral cloud into an \ion{H}{2} region. The observed (net) polarization spanning the Lacy jet (Figures 2 and 3) is only 1-2\% at 8.7 and 12.5 $\mu$m, which is a local minimum in the polarimetric images at those wavelengths; the observed 10.3 $\mu$m polarization of 3-4$\%$ across the same region appears to arise in foreground material. The Lacy jet is not identified with a unique polarimetric signature in our observations. 

\emph{W51d} was first detected at 2 and 6 cm by Wood \& Churchwell (\hyperlink{woo89}{1989}; see also Ginsburg et al. \hyperlink{gin16}{2016}), who showed it to be strongly peaked and extended on the scale of several arcsec. The radio emission arises in ionized gas filling a UC \ion{H}{2} region/cometary nebula. The primary peak in the radio emission was identified with a compact NIR source by Goldader \& Wynn-Williams (\hyperlink{gol94}{1994}). A massive star (shown as a star symbol in the figures), identified in K-band images by Figuerêdo et al. (\hyperlink{fig08}{2008}) and coincident with the NIR peak, is probably the most evolved star in IRS2, dissipating its surroundings enough for its photospheric features to be observable (Barbosa et al. \hyperlink{barbosa16}{2016}). The star is spectroscopically classified as approximately O3 (Barbosa et al. \hyperlink{barbosa08}{2008}) and likely drives that cometary nebula outflow.

\emph{IRS2W} has sometimes been conflated with the source W51d, leading to some confusion in the literature. Here we follow the identification used by Barbosa et al. (\hyperlink{barbosa16}{2016}) in which IRS2W coincides with the resolved and weaker secondary radio peak detected by Wood \& Churchwell (\hyperlink{woo89}{1989}) about 1$\arcsec$ SW of the W51d NIR source mentioned above. In this case, the SW radio emission peak is embedded in a broad, somewhat elongated peak or ridge in mid-IR emission (Okamoto et al. \hyperlink{oka01}{2001}; this work). Both the SW radio peak and the mid-IR peak/ridge appear to be on the boundary of the cometary nebula and are therefore likely powered by the star located at W51d, since IRS2W does not appear to coincide with a near-IR source. The observed polarization for IRS2W results from a combination of emissive and absorptive polarization, which near 10 $\mu$m, are comparable. We discuss IRS2W in more detail below.

{\subsection{B-Field Morphology: General Trends}} \label{sec:3.3}
               
We detect absorptive and emissive polarization magnitudes across W51 IRS2 that are among the highest ever detected. Until now, \emph{p}{\textsubscript{a}} $\approx$ 12.5\% and \emph{p}{\textsubscript{e}} $\approx$ 12.5\% from the Becklin-Neugebauer object in Orion (Smith et al. \hyperlink{smith00}{2000}; Aitken et al. \hyperlink{ait89}{1989}) and the Galactic Center core region (Roche et al. \hyperlink{Roc18}{2018}), respectively, have been the highest observed values. Aitken et al. (\hyperlink{ait98}{1998}) have argued that grain alignment in the Galactic Center is saturated and therefore the polarization there cannot increase above 12\%. Something similar may be occurring in W51 IRS2, which could indicate that this is a common occurrence for both polarized emission and absorption, a conclusion that could impose constraints on the grain alignment mechanism itself. 

Figure 4a shows that the projected field B\textsubscript{a} exhibits a preponderance of field lines on the right-hand side of the image with $\theta$\textsubscript{a} in the range 110\degr - 160\degr. That region overlaps W51d/IRS2W. Near the middle and left-hand side of the image, $\theta$\textsubscript{a} decreases, with some very low values of polarization and more patchiness evident across the distribution. The trend is most evident in the histogram for $\theta$\textsubscript{a} shown in Figure 5a (red line), which shows a very pronounced peak in the distribution near 140\degr.
        
\begin{figure}
\resizebox{\hsize}{!}{\includegraphics{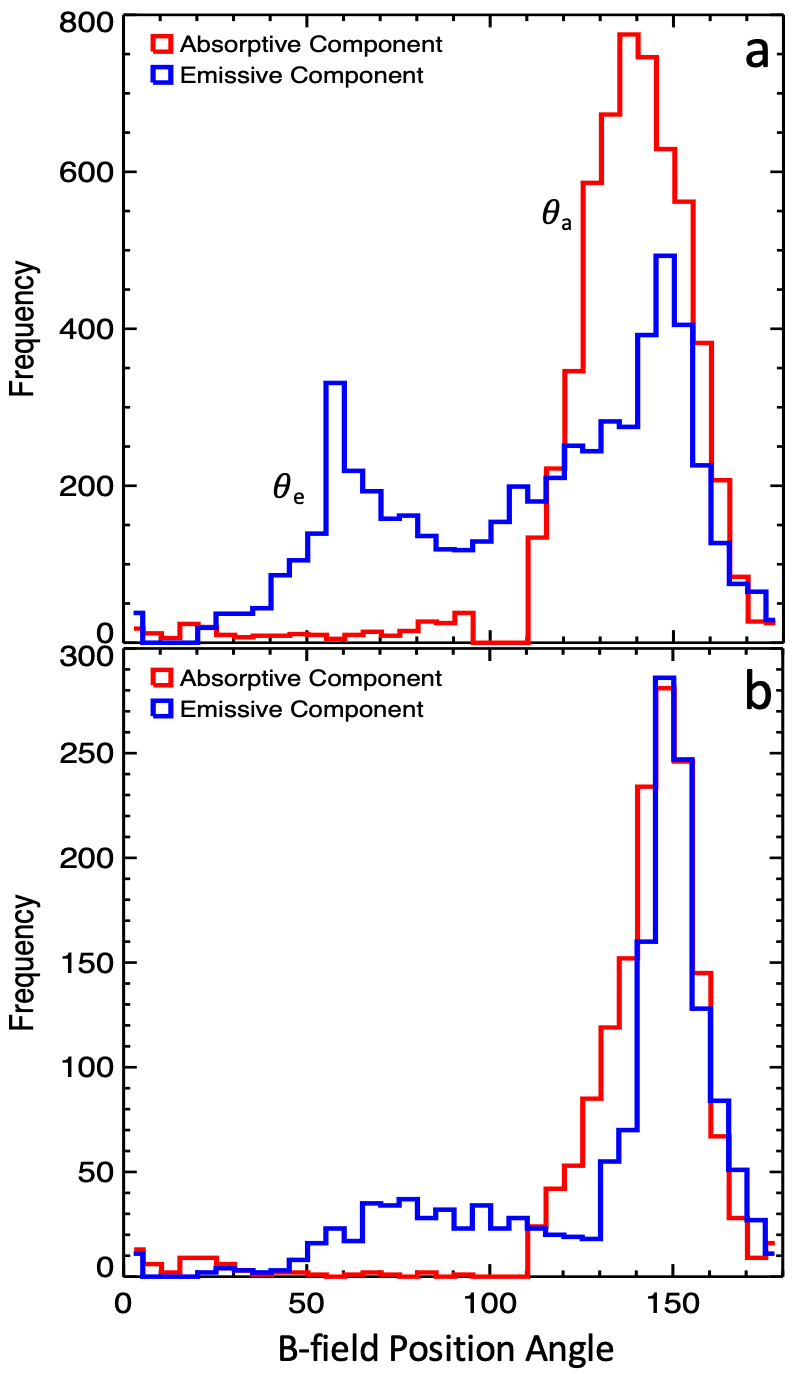}}
\caption{Histograms of position angles $\theta$\textsubscript{e} and $\theta$\textsubscript{a} of B-fields in the emitting and absorbing regions, respectively, and spanning: (a) the entire observed region, and (b) the champagne outflow sub-region extending horizontally from 1\farcs1 to 7\farcs6 and vertically from 1\farcs1 to 2\farcs7 in Figure 4. Each bin is 5\degr wide.
}\label{Figure5}
\end{figure}
        
In contrast, the derived polarization of the emitting region (Figure 4b) clearly exhibits much larger variations in both magnitude and direction across our field of view than does the absorbing region. The two regions exhibiting the highest fractional polarization and polarized emission (most evident as yellow contours in Figure 4b) are mutually orthogonal: the upper-right quadrant of our plot spanning IRS2W and W51d, and much of the upper left side of the plot centered near W51d1. The dominance of these two zones is evident in the marked bimodal distribution (blue line) for $\theta$\textsubscript{e} in Figure 5a. These two prominent regions are sharply demarcated by their higher polarization fractions and intensities, and they are separated from each other by a ``gulf" of much lower polarization, which extends in a 2\arcsec-wide swath from north to south and along much of the southern part of the plot; the polarization in those areas is $<$2$\%$ with generally ill-defined PA values, although greater smoothing of the data, as in Figure 6, suggests somewhat more coherence in the PA distribution there than is evident in Figure 4. 

Especially apparent in Figure 4b are the two high-polarization ridges ($>$10$\%$) coincident with W51d and IRS2W. Together these two ridges appear to define an (almost) C-shaped polarized arc with the projected B-fields aligned along the horns of the arc, as considered further in Section 4.2.1.The virtually identical values for $\theta$\textsubscript{a} and $\theta$\textsubscript{e} across much of IRS2W, evident in Figures 4 and 5b, likely reflect details of the champagne-outflow physics associated with the cometary nebula, as discussed below.     

Using a $\sim$5\arcsec aperture for their mid-IR spectropolarimetry, Smith et al. (\hyperlink{smith00}{2000}) observed a peak polarization near 10 $\mu$m of $\sim$4$\%$ at a PA of 142$\degr$. The exact location they observed in W51 IRS2 is uncertain, but their beam was likely centered somewhere within the complex of bright mid-IR sources W51d, OKYM2, IRS2E, and the Lacy jet, all of which could have been spanned by their $\sim$5\arcsec aperture (Figure 1; Barbosa et al. \hyperlink{barbosa16} {2016}). In this regard, we note that our observed spectropolarimetry and modeling using the Aitken method for the Lacy jet match in detail those presented by Smith et al. (\hyperlink{smith00}{2000}) for W51 IRS2. We conclude that our high-resolution observations are consistent with those by Smith et al. (\hyperlink{smith00}{2000}).\newline
        
\section{Discussion} \label{sec:4}

In this section we examine in more detail the properties of the mid-IR emitting and absorbing regions, which in our simple picture are threaded by B-fields with projected components B\textsubscript{e} and  B\textsubscript{a}, respectively. We first focus on the relationship of outflowing ionized gas from the IRS2W cometary nebula to the likely co-spatial B-field, after which we consider the possible relationship between the B-field and possible large-scale accretion.  
   
    {\subsection{B-Field in Absorbing Region: the Ambient Field}} \label{sec:4.1}

Except at IRS2W, the B\textsubscript{a} morphology does not appear correlated with any of the sources within the W51 IRS2 region, suggesting that it is not intimately associated with most mid-IR-emitting sources there. Instead, the overall morphology of B\textsubscript{a} resembles that of the lower resolution ($\sim$20\arcsec, or 0.5 pc) Caltech Submillimeter Observatory (CSO) polarimetry at 350 $\mu$m (Dotson et al. \hyperlink{dot10}{2010}), which spans a much larger region including that considered here; seen in emission, the observed PA values of the 350 $\mu$m polarization must be rotated by 90\degr to derive the corresponding B-field direction, which is consistent with our inferred B\textsubscript{a}. Tang et al. (\hyperlink{tang13}{2013}) conclude that the CSO polarimetry and 850 $\mu$m polarimetry by Chrysostomou et al. (\hyperlink{chr02}{2002}) trace emission in foreground molecular cloud material. Based on this comparison, we identify B\textsubscript{a} with that foreground molecular material, which may represent the larger-scale ambient field.

{\subsection{B-Field across IRS2W Emitting Region}} \label{sec:4.2}

{\subsubsection{Relationship to Ionized Gas}} \label{sec:4.2.1}

Radio continuum and recombination-line observations with $\sim$0\farcs4 resolution by Ginsburg et al. (\hyperlink{gin16}{2016}) show that the compact cometary nebula IRS2W is the source of complex gas outflow. In Figure 6 we overlay the polarization line segments for the emitting region onto part of their 14.5 GHz continuum image; these line segments differ from those in Figures 2 and 4 by resulting from 5 $\times$ 5 pix, (0\farcs4 $\times$ 0\farcs4) binning to lessen crowding in the figure. Prominent tongue-like features of ionized-gas emission fan into the north and northwest from the limb-brightened cometary core. H77$\alpha$ observations indicate that gas along our sightline to the cometary core is blue-shifted by 25-30 km s\textsuperscript{-1} relative to the core and almost certainly emanating from it (Ginsburg et al. \hyperlink{gin16}{2016}). If the blue-shifted material and fan-like outflows are different components of the same approaching complex, then the outflow velocities are of order 35 km s\textsuperscript{-1}.
             
\begin{figure*}[t]
\centering\includegraphics[width=\textwidth]{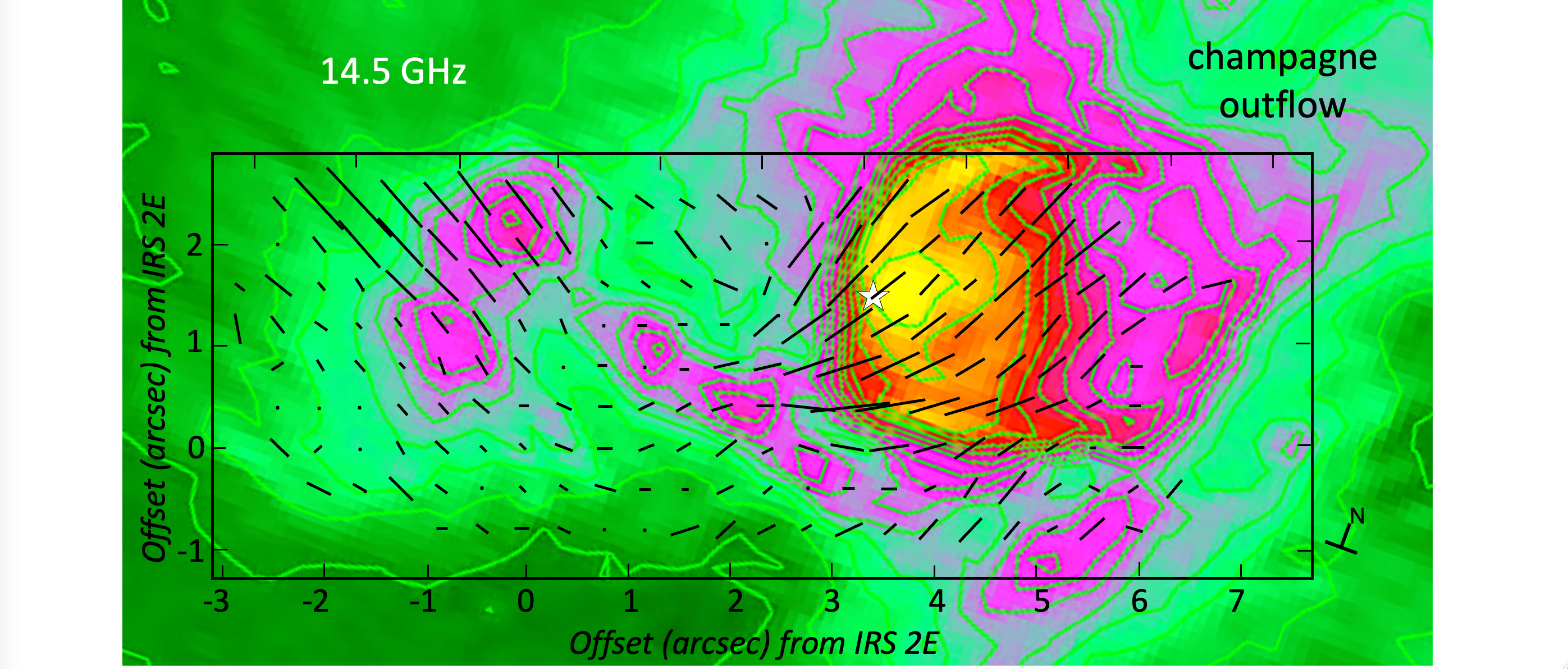}
\caption{
    Overlay of W51-IRS2 emissive-polarization line segments on 14.5 GHz VLA continuum image by Ginsburg et al. (\protect\hyperlink{gin16}{2016}). The line segments are rotated by 90\degr to indicate the orientation of B\textsubscript{e}, as in Figure 4. The star symbol indicates the location of the exciting star of the IRS2W cometary nebula (Figure 1). The champagne outflow discussed in the text is also indicated. To avoid crowding, polarization data points have been binned by 5 $\times$ 5 pix, (0\farcs4 $\times$ 0\farcs4), in contrast to lower binning for plots in Figures 2 and 4.
}\label{Figure6}
\end{figure*}
             
The polarization line segments, and therefore the projected B\textsubscript{e} field lines,  across the core of IRS2W intersect the outflow features. Interestingly, the symmetry axis of the C-shaped cometary head is oriented along the horizontal axis of the plot in Figure 6, with the horns of the C pointing to the plot's right. If we accept that the horns define the projected opening from the UC \ion{H}{2} region through which the ionized gas is flowing, then, in projection, the gas outflow and the B-field are exiting the opening at a projected angle of $\sim$45\degr with respect to the nebular symmetry axis. By referring to the less-binned polarimetric data in Figure 4b, we see that the C-shaped ridge prominent at 14.5 GHz has nearly twice the polarized intensity and fractional polarization than the immediately adjacent surroundings.  The projected B-field is not only higher in that ridge but seems to follow its curvature.  
           
Based on this comparison of projected morphologies, we conclude that the B-field and outflowing gas are physically coupled and tilted significantly relative to the direction expected for unperturbed, newly ionized gas flowing away from an ionization front. In addition, it appears that the B$\textsubscript{e}$ field lines at the south ionized wall, or horn, of the C-shaped cometary head may be nearly parallel to the wall. If this is correct, it would suggest compression of the B-field there as the high-pressure outflowing gas is constrained and funnelled through the opening.    
           
{\subsubsection{Outflow Energetics}} \label{sec:4.2.2}
   
If, as implied by their mutual alignment, the outflowing gas and emitting-region B-field associated with IRS2W are coupled, what can we say about the relative importance of the B-field and gas in determining the observed morphology in the cometary nebula and its immediate vicinity? The relevant physical quantities are the kinetic energy density U\textsubscript{KE} of the outflowing gas, the energy density U\textsubscript{B} of the co-spatial B-field, and the balance between them represented by the parameter $\kappa$:
    
\begin{equation}
            \kappa = \frac{U\textsubscript{KE}}{U\textsubscript{B}} = {\frac{4\pi \rho v\textsuperscript{2}}{B\textsuperscript{2}}}, 
\end{equation}
where $\rho$ is the gas (proton) density, \emph{v} the outflow speed, and \emph{B} the B-field strength.
       
No direct measurements of the B-field strength exist for the W51 IRS2 region. However, by using 2 mm emission (Zhang et al. \hyperlink{zhang98}{1998}) to derive the mass of the  $\sim$9\arcsec -diameter (0.24 pc, adjusted to a distance of 5.4 kpc) region spanning W51 IRS2, and applying the virial theorem, Chrysostomou et al. (\hyperlink{chr02}{2002}) estimate an upper limit of $\sim$0.6 mG for the B-field across this region. Using the Chandrasekhar-Fermi technique, Lai et al. (\hyperlink{lai01}{2001}) estimate lower limits of 0.2 - 1.3 mG, depending on assumed gas densities, for the strength of the projected parsec-scale B-field in the vicinity of W51 e1/e2, which is $\sim$2 pc southeast of W51 IRS2; Chrysostomou et al. (\hyperlink{chr02}{2002}) estimate an upper limit of 3.5 mG for that same region using the same data of Lai et al. These considerations suggest that assuming a value of 1 mG for the B-field in the vicinity of IRS2W, which has a diameter of $\sim$0.07 pc, is reasonable.
   
To estimate the gas density near the base of the outflow, i.e., near the location of gas egress from the cometary \ion{H}{2} region IRS2W itself, we note that the intensity of the H77$\alpha$ recombination line near that location, as measured with 0\farcs4 resolution by Ginsburg et al. (\hyperlink{gin16}{2016}, their Figure 7), is about a factor of two greater than the intensity they measure (their Figure 5) at the bright shell-like feature located in the W51 Main/IRS1 region. There they derive a value for the emission measure of $\sim$10\textsuperscript{7} cm\textsuperscript{-6} pc for a 0.03-pc path length appropriate for their source and a corresponding electron density of 2.4 x 10\textsuperscript{4} cm\textsuperscript{-3}. Based on its projected size of 0.07 pc and relative H77$\alpha$ intensity, we then estimate an electron density of 1.6 x 10\textsuperscript{4} cm\textsuperscript{-3} at the base of the IRS2W outflow. Therefore, for a comparable proton density, the outflow mass density there is $\rho$ = 3 x 10\textsuperscript{-20} g cm\textsuperscript{-3}.
  
Assuming this value of the mass density, an outflow speed of 35 km s\textsuperscript{-1}, and a 1 mG B-field, we find that $\kappa$ = 5 in the vicinity of IRS2W. For a 0.6 mG field, $\kappa$ = 14. It appears that the gas outflow dominates energetically and  can potentially distort significantly the B-field in the immediate vicinity of the IRS2W cometary nebula at this stage of its evolution. In fact, $\kappa >$ 1 for B $<$ 2 mG, so that even for higher credible values of B-field strength the outflowing gas must have a significant impact on the B-field configuration. 

\vspace{.1in}
      
\begin{figure*}[t]
\centering\includegraphics[width=\textwidth]{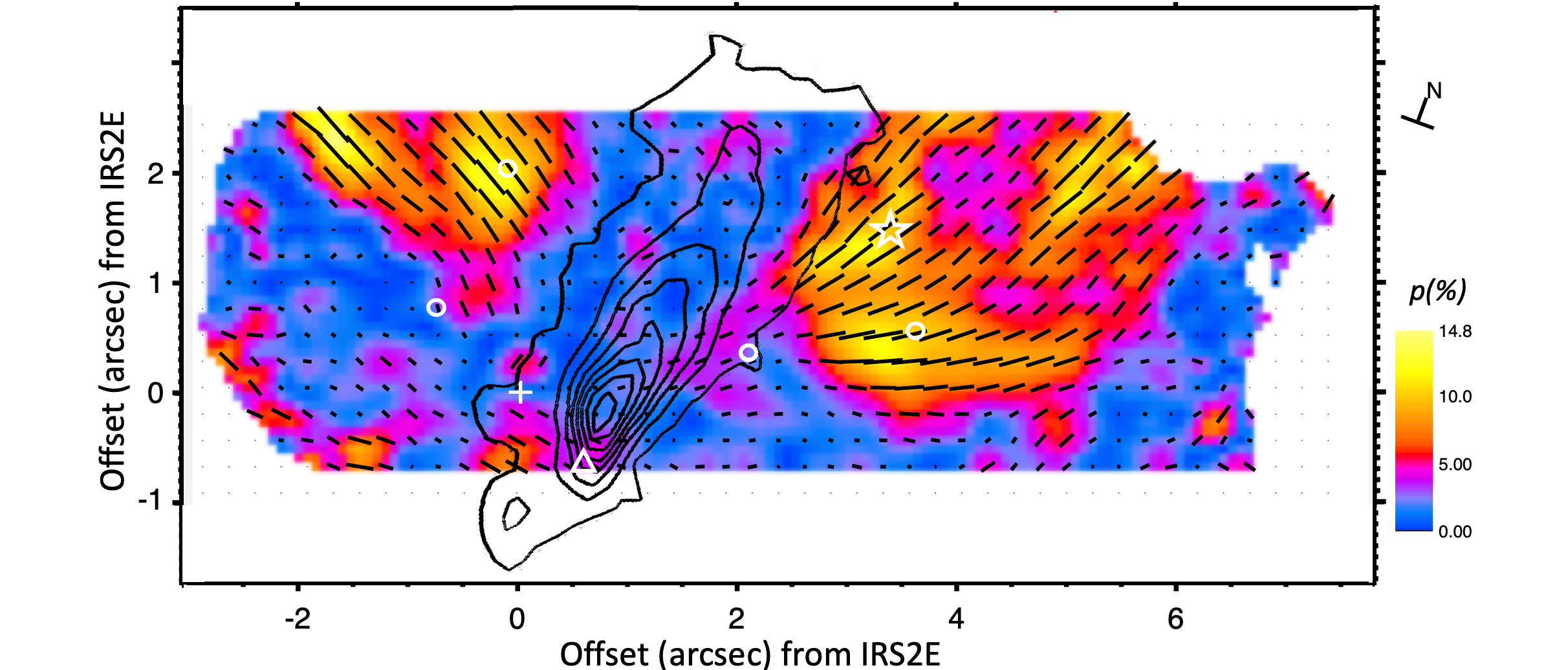}
\caption{
    \textsuperscript{12}CO J=2-1 line emission (black contours), associated with jet outflow originating near center of our field of view (Goddi et al. \hyperlink{god20}{2020}), overlaid on polarization distribution corresponding to B\textsubscript{e} (same as in Fig. 4b). All CO line emission north of SMA1/N1 is red-shifted, with blue-shifted CO emission extending southward from very near SMA1/N1 (see Goddi et al. \hyperlink{god20}{2020}). CO contours are linearly spaced with values 6, 17, 29, 41, 53, 64, 76, and 88\% the peak value.
}\label{Figure7}
\end{figure*}

{\subsubsection{Champagne Outflow from Cometary Nebula}} \label{sec:4.2.3}
The B\textsubscript{e} field lines across IRS2W extend from the cometary nebula toward the ionized-gas outflow features; the origin of the outflows is identified with the IRS2W cometary nebula, and the outflowing gas appears to be a good example of a so-called ``champagne outflow." Hydrodynamic and magnetohydrodynamic (MHD) simulations show that a young massive star embedded near the edge of a molecular cloud can produce a cometary \ion{H}{2} region with a prominent gas outflow tail that has broken through the molecular cloud surface. Limb brightening on the higher-density molecular-cloud side of the \ion{H}{2} region is a defining characteristic of the cometary head. When there is no B-field, the gas outflow, or ``champagne flow," erupts from the molecular cloud in a direction opposite the cometary head (Steggles et al. \hyperlink{steg17}{2017}; Arthur \& Hoare \hyperlink{art06}{2006}; Gendelev \& Krumholz \hyperlink{gen12}{2012}, who refer to this as a blister-type \ion{H}{2} region). However, a B-field threading the region can influence the dynamics of the outflowing gas.  

Qualitatively, our results for the champagne outflow from IRS2W resemble those of the MHD simulations by Gendelev \& Krumholz (\hyperlink{gen12}{2012}) for the outflows in magnetized gas clouds, although they incorporate a very different regime of density, temperature, ionization, and B-field strength than those applicable to W51 IRS2. Most relevant to our study is the simulation with the B-field oriented at 45\degr to the molecular cloud surface. Gendelev \& Krumholz show that the ionized gas is initially propelled outward and away from the ionization front in the cometary head and perpendicular to the molecular cloud surface. However, the outflow becomes progressively more constrained and aligned with the large scale B-field as it progresses. As discussed above, we have proposed that the B$\textsubscript{a}$ distribution derived from our polarimetry corresponds to the larger scale ambient B-field. Comparison of the B-field distributions in Figures 4a and 4b shows that $\theta$$\textsubscript{e}$ and $\theta$$\textsubscript{a}$ are comparable in value (within 20\degr) for those regions coinciding almost exclusively with IRS2W, in keeping with expectations based on the MHD champagne-outflow simulations.\\
\newline
{\subsection{Relationship of B-Field to W51north Jet and Accretion Flow}}\label{sec:4.3}
     
The brightest 230 GHz (1.3 mm) source in W51 IRS2 is SMA1 (Tang et al. \hyperlink{tang13}{2013}), which appears to coincide with the bright 870 $\mu$m source N1 (Koch et al. \hyperlink{koch18}{2018}). SMA1/N1 is thought to be surrounded by a large, massive circumstellar dust disk and an even larger infalling and rotating molecular gas ring, or toroid. SMA1/N1 is also the likely origin of a remarkable bipolar outflow extending $\sim$20,000 AU to the north (Zapata et al. \hyperlink{zap09}{2009}, \hyperlink{zap10}{2010}; Goddi et al. \hyperlink{god20}{2020}); following Goddi et al. (\hyperlink{god20}{2020}), we refer to this outflow as the W51north jet. In Figure 7 we show an overlay of the W51north jet CO line emission onto our image of the polarized emission, which traces B\textsubscript{e}. The jet converges on SMA1/N1 and coincides almost exactly with the clearly demarcated north-south swath of significantly smaller B\textsubscript{e}. While not so obvious in Figure 4b, the small projected B-field within that swath may, in fact, have a relatively smooth, coherent morphology; the more-smoothed data plotted in Figure 6 reveals a relatively continuous transition in $\theta$$\textsubscript{e}$ from east to west.
 
Using their 0$\farcs$7-resolution polarimetry at 870 $\mu$m, Tang et al. (\hyperlink{tang13}{2013}) note that the inferred B-field morphology spanning a $\sim$2\arcsec diameter region roughly centered on SMA1/N1 is consistent with one that can channel gas to the dense core. In Figure 8 we overlay their (rotated) polarization line segments onto a portion of our B\textsubscript{e} image centered near SMA1/N1. Inspection of Figure 8 indicates that within 1\arcsec of SMA1/N1, there is no obvious correlation between the mid-IR and 870 $\mu$m B-field angles; in fact, they may be orthogonal to each other there. However, if the vertical group of 870 $\mu$m line segments closest to SMA1/N1 is extrapolated toward W51d1, they can plausibly connect along an arc to those near W51d1. 

While speculative, such a connection could support a picture in which SMA1/N1 is a major, even dominant, determinant of B-field morphology across W51 IRS2 while also accounting for other observed properties of this region such as collimation of the W51north jet (e.g., Reissl et al. \hypertarget{rei17}{2017}). The 870 $\mu$m line group west of SMA1/N1 is oriented roughly parallel to both the B\textsubscript{e} and B\textsubscript{a} field lines spanning the IRS2W region, and it remains unclear how it fits into a larger picture of the region that includes our mid-IR polarimetry. \\

\begin{figure}[h]
\resizebox{\hsize}{!}{\includegraphics{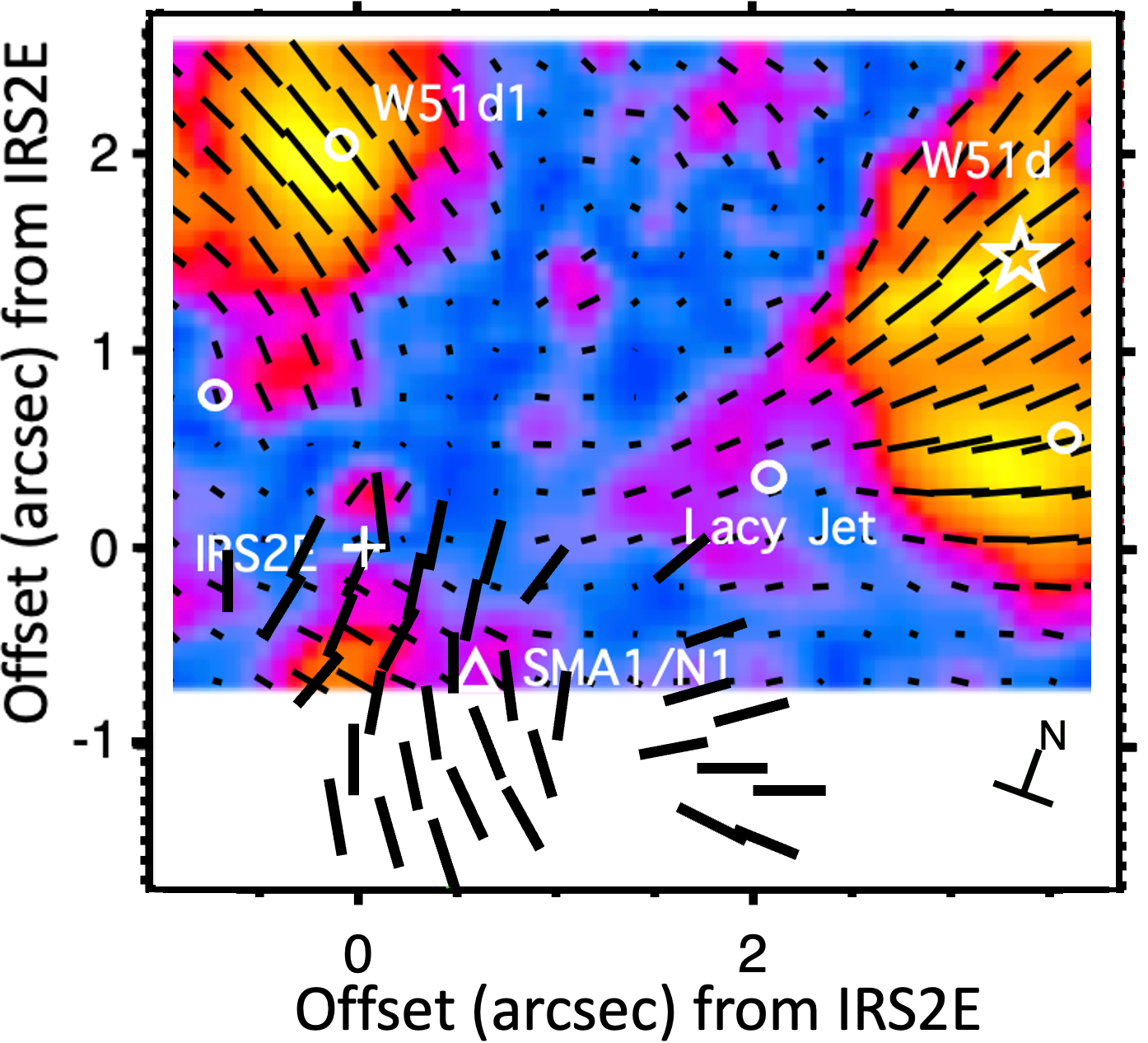}}
\caption{
    Overlay of B-field (thick black line segments), obtained with 0$\farcs$7-resolution polarimetry at 870 $\mu$m by Tang et al. (\protect\hyperlink{tang13}{2013}), onto a portion of our B\textsubscript{e} image centered near SMA1/N1 (triangle).
}\label{Figure8}
\end{figure}
\vfill 
\section{Conclusions} \label{sec:5}

We have used the technique developed by Aitken et al. (\hyperlink{ait04}{2004}) to unfold our mid-IR measurements of the polarisation of W51 IRS2 into two components, one due to warm, emitting silicate dust grains and the other due to absorption by colder foreground silicate grains. We are thereby able to infer properties of the respective B-fields, B\textsubscript{e} and B\textsubscript{a}, in the plane of the sky and along each sightline: in effect, tomography. The principal conclusions of this work are as follows:

    1. Based on comparison of our data with that at mm and submm wavelengths, we identify B\textsubscript{a} with foreground molecular material and representative of the larger-scale ambient field. 
    
    2. The morphologies of the projected B-fields in the mid-IR-emitting and foreground-absorbing regions spanning the cometary \ion{H}{2} region W51 IRS2W are similar. Elsewhere, however, the two fields differ significantly with no clear relationship between them.
    
    3. The projected B-field spanning IRS2W is likely an integral part of a champagne outflow originating in the cometary \ion{H}{2} region, whereby outflowing ionized gas is funneled along a pre-existing ambient B-field. The kinetic energy density of that gas is more than a few times the B-field energy density, and therefore the outflowing gas probably significantly modifies the B-field morphology near the outflow base.
    
    4. The remarkable bipolar outflow, or W51north jet, that appears to originate at or near SMA1/N1 coincides almost exactly with the clearly demarcated north-south swath of smaller B\textsubscript{e}. That swath separates the two regions of W51 IRS2 that have B-field orientations nearly orthogonal to each other. The (much smaller) B-field lines within that swath seem to smoothly connect to the B-fields in those two regions.
    
    5. Some, but not all, of the B-field morphology close to SMA1/N1 and determined from submm observations could plausibly connect smoothly to the larger scale structure that we see in the W51d1 region. That could support a picture in which SMA1/N1 plays a major role in the B-field structure across W51 IRS2.

\newpage
\section*{Appendix} \label{sec:appendix}

In general, polarization in the mid-IR results from a combination of dichroic absorption and emission by aligned non-spherical dust grains. Since, for the same grain alignment, there is a 90\degr difference in the polarization PA for absorption and emission, we must separate them in order to derive the B-field morphology along the sightline. Aitken et al. (\hyperlink{ait04}{2004}) developed a strategy to do this. 

Their approach relies on the fact that silicate dust emissivity exhibits a strong, fairly broad spectral feature spanning the 8-14 $\mu$m wavelength region. While there are variations in the shape of this feature from region to region, a large fraction of the YSOs studied spectrophotometrically by Smith et al. (\hyperlink{smith00}{2000}) exhibit a feature very similar to that emitted from hot dust in Orion. Equations 2 and 3, which are wavelength dependent, are the expressions for absorptive polarization $p\textsubscript{a}$ and emissive polarization $p\textsubscript{e}$ as a function of $\tau_{\parallel}$, the optical depth for radiation with electric vector parallel with the dust-particle long axis, and $\tau_{\perp}$, the optical depth for radiation with electric vector perpendicular to the long axis; it is the (usually small) difference $\tau_{\parallel}$ - $\tau_{\perp}$ that gives rise to the observed polarization.  

\begin{equation}
    p_{\text{a}} = \frac{e^{-\tau\parallel}-e^{-\tau\perp}}{{e^{-\tau\parallel}}+e^{-\tau\perp}} = -\frac{{\text{tanh}[(\tau_\parallel-\tau_\perp)]}}{2}\approx -\frac{(\tau_\parallel-\tau_\perp)}{2}
\end{equation}

\begin{equation}
    p_{\text{e}} = \frac{(1-e^{-\tau\parallel})-(1-e^{-\tau\perp})}{(1-e^{-\tau\parallel})+(1-e^{-\tau\perp})}\approx\frac{\tau_\parallel-\tau_\perp}{\tau_\parallel+\tau_\perp}
\end{equation}
\\
Thus, we have $p\textsubscript{e} \approx  -p\textsubscript{a}/\tau$, where $\tau \approx (\tau_{\parallel} + \tau_{\perp}$)/2. The approximation in Equation 3 arises from the fact that observed mid-IR emitting regions are usually optically thin. On the other hand, foreground absorbing regions can be optically thick, with the magnitude of polarization being proportional to the optical depth difference, as noted.    

Aitken et al. (\hyperlink{ait04}{2004}) assume that the observed spectropolarimetry of the BN object in Orion is purely absorptive. They then use $\tau$ from the observation of Trapezium region of Orion to derive the pure emissive polarization profile as \emph{f}$\textsubscript{e}$($\lambda$) = \emph{f}$\textsubscript{a}$($\lambda$)/$\tau$($\lambda$). The observed normalized Stokes parameters \emph{q} and \emph{u} are assumed to be linear combinations of Stokes parameters arising from dichroic absorption and emission as shown in the following relations, where \emph{f}$\textsubscript{a}$($\lambda$) and \emph{f}$\textsubscript{e}$($\lambda$)are the absorption and emission polarization profiles, respectively, normalized to unity at the profile peak: 10.2 $\mu$m for absorption and 11.5 $\mu$m for emission. 

\begin{equation}
    q(\lambda) = q_{\text{a}}(\lambda) + q_{\text{e}}(\lambda) = Af_{\text{a}}(\lambda)+ Bf_{\text{e}}(\lambda)
\end{equation}

\begin{equation}
    u(\lambda) = u_{\text{a}}(\lambda) + u_{\text{e}}(\lambda) = Cf_{\text{a}}(\lambda) + Df_{\text{e}}(\lambda)
\end{equation}

As discussed in detail in Aitken et al. (\hyperlink{ait04}{2004}), the coefficients \emph{A} and \emph{C} are the fitted values of \emph{q}$\textsubscript{a}$ and \emph{u}$\textsubscript{a}$ at 10.2 $\mu$m, and   \emph{B} and \emph{D} are the fitted values of \emph{q}$\textsubscript{e}$ and \emph{u}$\textsubscript{e}$ at 11.5 $\mu$m. The associated profile functions \emph{f}($\lambda$) give the values at other wavelengths. Values of \emph{p}$\textsubscript{a}$ or \emph{p}$\textsubscript{e}$ and their corresponding position angles are then derived directly from \emph{A} and \emph{C} or \emph{B} and \emph{D}, respectively. Once we have the polarization measurements for at least two wavelengths, it is feasible to separate absorption and emission polarization components. However, as noted in the text, we have used three passbands, one centered on the central part of the silicate feature and the other two on the short- and long-wavelength sides.

\section*{Acknowledgements} \label{sec:acknowledgements}
This research is based on observations made with the Gran Telescopio CANARIAS (GTC), installed at the Spanish Observatorio del Roque de los Muchachos of the Instituto de Astrofísica de Canarias, on the island of La Palma. It is a pleasure to acknowledge the outstanding support of the GTC science and engineering staff who made these observations possible. We also thank Dr. Adam Ginsburg for useful comments and insights. This research was supported in part by the National Science Foundation under grant AST-1908625 to CMT. CMW acknowledges financial support during the period 2012-2017 from an Australian Research Council Future Fellowship FT100100495. The upgrade of CanariCam was co-financed by the European Regional Development Fund (ERDF), within the framework of the ``Programa Operativo de Crecimiento Inteligente 2014-2020", project ``Mejora de la ICTS Gran Telescopio CANARIAS (2016-2020)."

\newpage
\section*{References} \label{sec:refs}
    \hypertarget{ait04}
        {Aitken, D. K., Hough, J. H., Roche, P. F., Smith, C. H., \& Wright, C. M. 2004, MNRAS, 348, 279} 

    \hypertarget{ait89}
        {Aitken, D. K., Smith, C. H., \& Roche, P/ F. 1989, MNRAS, 236, 919} 

    \hypertarget{ait98}
        {Aitken, D. K., Smith, C. H., Moore, T. J. T., \& Roche, P/ F. 1998, MNRAS, 299, 743} 
        
    \hypertarget{art06}
        {Arthur, S. J., \& Hoare, M. G. 2006, ApJS, 165, 283} 
        
    \hypertarget{barbosa08}
        {Barbosa, C. L., Blum, R. D., Conti, P. S., Damineli, A., \& Figuerêdo, E. 2008, ApJL, 678, L55} 
        
    \hypertarget{barbosa16}
        {Barbosa, C. L., Blum, R. D., Damineli, A., Conti, P. S., \& Gusmão, D. M. 2016, ApJ, 825, 54} 
        
    \hypertarget{barnes15}
        {Barnes, P., Li, D., Telesco, C., et al. 2015, MNRAS, 453, 2622} 
        
    \hypertarget{cho07}
        {Cho, J., \& Lazarian, A. 2007, ApJ, 669, 1085} 
    
    \hypertarget{chr02}    
        {Chrysostomou, A., Aitken, D. K., Jenness, T., et al. 2002, A\&A, 385, 1014} 
    
    \hypertarget{cohen99}
        {Cohen, M., Walker, R. G., Carter, B., et al. 1999, AJ, 117, 1864} 
        
    \hypertarget{dot10}
        {Dotson, J. L., Vaillancourt, J.E., Kirby, L., et al. 2010, ApJS, 186, 406} 

    \hypertarget{dra21}
        {Draine, B. T., \& Hensley, B. S. 2021, ApJ, 920, 47} 
        
    \hypertarget{esw17}
        {Eswaraiah, C., Lai, S.-P., Chen, W.-P., et al. 2017, ApJ, 850, 195} 
    
    \hypertarget{fig08}
        {Figuerêdo, E., Blum, R. D., Damineli, A., Conti, P. S., \& Barbosa, C. L. 2008, AJ, 136, 221} 
    
    \hypertarget{gen12}
        {Gendelev, L., \& Krumholz, M. R. 2012, ApJ, 745, 158} 
    
    \hypertarget{gin15}
        {Ginsburg, A., Bally, J., Battersby, C., et al. 2015, A\&A, 573, A106} 
    
    \hypertarget{gin16}
        {Ginsburg, A., Goss, W. M., Goddi, C., et al. 2016, A\&A, 595, A27} 

    \hypertarget{gin17}
        {Ginsburg, A., Goddi, C., Kruijssen, J. M. D., et al. 2017, ApJ, 842, 92} 
        
    \hypertarget{god20}    
        {Goddi, C., Ginsburg, A., Maud, L., Zhang, Q., \& Zapata, L. 2020, ApJ, 905, 25}
        
    \hypertarget{gol94}
        {Goldader, J. D., \& Wynn-Williams, C. G., 1994, ApJ, 433, 164} 
        
    \hypertarget{hil88}
        {Hildebrand, R. H., 1988, QJR astr.Soc, 29, 327} 
    
    \hypertarget{hull17}
        {Hull, C. L. H., Girart, J. M., Tychoniec, L., et al. 2017, ApJ, 847, 92} 
    
    \hypertarget{kat15}
        {Kataoka, A., Muto, T., Momose, M., et al. 2015, ApJ, 809, 78} 
    
    \hypertarget{kat16}
        {Kataoka, A., Tsukagoshi, T., Momose, M., et al. 2016, ApJL, 831, L12} 
     
    \hypertarget{koch18}
        {Koch, P. M., Tang, Y.-W., Ho, P. T. P., et al. 2018, ApJ, 855, 39} 
        
    \hypertarget{krum19}
        {Krumholz, M. R., \& Federrath, C. 2019, Frontiers in Astro \& Sp. Sc., 6, 1. doi: 10.3389/fspas.2019.00007}  
    \hypertarget{krum07}
        Krumholz, M. R., Stone, J. M., \& Gardiner, T. A., 2007, ApJ, 671, 518. doi: 10.1086/522665
    
    \hypertarget{lacy07}
        {Lacy, J. H., Jaffe, D. T., Zhu, Q., et al. 2007, ApJL, 658, L45} 
    
    \hypertarget{lai01}
        {Lai, S.-P., Crutcher, R. M., Girart, J. M., \& Rao, R. 2001, ApJ, 561, 864} 
    
    \hypertarget{laz07}
        {Lazarian, A. 2007, JQRST, 106, 225} 
    
    \hypertarget{li14}
        {Li, D. 2014, iDealCam: Interactive Data Reduction and Analysis for
        CanariCam, Astrophysics Source Code Library, ascl:1411.009} 
    
    \hypertarget{li18}
        {Li, D., Telesco, C. M., Zhang, H., et al. 2018, MNRAS, 473, 1427} 
    
    \hypertarget{loro16}
        {Lopez-Rodriguez, E. 2016, MNRAS, 455, 2656} 
    
    \hypertarget{loro17}
        {Lopez-Rodriguez, E., Packham, C., Jones, T. J., et al. 2017, MNRAS, 464, 1762} 
            
    \hypertarget{mat09}
        {Matthews, B. C., McPhee, C.A., Fissel, L. M., \& Curran, R. L. 2009, ApJS, 182, 143}   
    
    \hypertarget{oka01}
        {Okamoto, Y. K., Kataza, H., Yamashita, T., Miyata, T., \& Onaka, T.
        2001, The Astrophysical Journal, 553, 254} 

    \hypertarget{pla14}
        {Plaszczynski, S., Montier, L., Levrier, F., \& Tristram, M. 2014 MNRAS, 439, 4048}

    \hypertarget{rei17}
        {Reissl, S., Seifried, D., Wolf, S., Banerjee, R., \& Klessen, R. S. 2017, A\&A, 603, A71} 

    \hypertarget{Roc18}
        {Roche, P. F., Lopez-Rodriguez, E., Telesco, C. M., Schödel, R., \& Packham, C. 2018, MNRAS, 476, 235} 
    
    \hypertarget{smith00}
        {Smith, C. H., Wright, C. M., Aitken, D. K., Roche, P. F., \& Hough, J. H. 2000, MNRAS, 312, 327} 

    \hypertarget{steg17}
        {Steggles, H. G., Hoare, M. G., \& Pittard, J. M. 2017, MNRAS, 466, 4573} 
    
    \hypertarget{steph17}
        {Stephens, I. W., Yang, H., Li, Z.-Y., et al. 2017, ApJ, 851, 55} 
    
    \hypertarget{tang13}
        {Tang, Y.-W., Ho, P. T. P., Koch, P. M., Guilloteau, S., \& Dutrey, A. 2013, ApJ, 763, 135} 
    
    \hypertarget{taz17}
        {Tazaki, R., Lazarian, A., \& Nomura, H. 2017, ApJ, 839, 56} 
    
    \hypertarget{telesco03}
        {Telesco, C. M., Ciardi, D., French, J., et al. 2003, in Society of
        Photo-Optical Instrumentation Engineers (SPIE) Conference Series, Vol. 4841, Instrument Design and Performance for Optical/Infrared
        Ground-based Telescopes, ed. M. Iye \& A. F. M. Moorwood, 913--922} 
    
    \hypertarget{woo89}
        {Wood, D. O. S., \& Churchwell, E. 1989, ApJS, 69, 831} 
    
    \hypertarget{xu09}
        {Xu, Y., Reid, M. J., Menten, K. M., et al. 2009, ApJ, 693, 413} 
    
    \hypertarget{yang16}
        {Yang, H., Li, Z.-Y., Looney, L., \& Stephens, I. 2016, MNRAS, 456, 2794} 
    
    \hypertarget{zap09}
        {Zapata, L. A., Ho, P. T. P., Schilke, P., et al. 2009, ApJ, 698, 1422} 
        
     \hypertarget{zap10}    
        {Zapata, L. A., Tang, Y.-W., \& Leurini, S. 2010, ApJ, 725, 1091}
        
    \hypertarget{zhang98}
        {Zhang, Q., Ho, P. T. P., \& Ohashi, H. 1998, ApJ, 494, 636} 
    
    \hypertarget{zhu15}
        {Zhu, Z. 2015, ApJ, 799, 16} 
    
\end{document}